\documentclass[aps, prb, twocolumn,floatfix,superscriptaddress,longbibliography, nofootinbib, sortedaddress]{revtex4-2}
\usepackage{graphicx, color, soul}
\usepackage{amsmath, amsfonts, amssymb, mathrsfs, dsfont, mathtools}
\usepackage[usenames,dvipsnames, pdftex, table]{xcolor}
\usepackage[normalem]{ulem} % needed for \scrap-command
\usepackage[mathscr]{eucal}
\usepackage{bm}
\usepackage{tabularx, booktabs}
\usepackage{multirow}
\usepackage[colorlinks, breaklinks, 
            linkcolor=NavyBlue,
            citecolor=NavyBlue,
            urlcolor=NavyBlue]{hyperref}
            
\usepackage[all]{hypcap} 
\usepackage{enumitem}

\newcommand{\Le}{\left}
\newcommand{\Ri}{\right}
\newcommand{\nn}{\nonumber}
\newcommand{\f}{\frac}
\newcommand{\bs}{\boldsymbol}

\newcommand{\ra}{\rangle}
\newcommand{\la}{\langle}
\newcommand{\eq}[1]{\begin{align}#1\end{align}}

\newcommand{\msr}{\mathscr}

\newcommand{\meang}{\overline{g}}
\newcommand{\gcorr}{\overline{g}_\mathrm{corr.}}
\newcommand{\nuI}{\nu_\mathrm{IQH}}
\newcommand{\ipr}{\overline{P_q}}
\newcommand{\Mc}{M_\mathrm{c}}
\newcommand{\br}{{\bm r}}

\date{\today}

\begin{document}
\title{Quantum Hall criticality in an amorphous Chern insulator}

\author{Soumya Bera}
\affiliation{Department of Physics, Indian Institute of Technology Bombay, Mumbai 400076, India}

\author{Johannes Dieplinger}
\affiliation{Institute of Theoretical Physics, University of Regensburg, D-93040 Germany}

\author{Naba P Nayak}
\affiliation{Department of Physics, Indian Institute of Technology Bombay, Mumbai 400076, India}

\begin{abstract}
We explore the critical properties of a topological transition in a two-dimensional, amorphous lattice with randomly distributed points. 
The model intrinsically breaks the time-reversal symmetry without an external magnetic field, akin to a Chern insulator. 
Here, the topological transition is induced by varying the density of lattice points or adjusting the mass parameter.
Using the two-terminal conductance and multifractality of the wavefunction, we found that the topological transition belongs to the same universality class as the integer quantum Hall transition.
Regardless of the approach to the critical point across the phase boundary, the localization length exponent remains within $\nu \approx 2.55-2.61$. 
The irrelevant scaling exponent for both the observables is $y\approx 0.3(1)$, comparable to the values obtained using transfer matrix analysis in the Chalker-Coddigton network.
Additionally, the investigation of the entire distribution function of the inverse participation ratio at the critical point shows possible deviations from the parabolic multifractal spectrum at the anomalous quantum Hall transition.
\end{abstract}
\maketitle

\section{Introduction}
The discovery of the integer quantum Hall effect~(IQH), characterized by the quantization of resistance and dissipationless chiral edge transport in a two-dimensional~(2D) electron gas under an external magnetic field, provides a fundamental understanding of the topological phenomena in condensed matter physics~\cite{KlitzingPRL80}. 
The quantum anomalous Hall effect~(QAH), as elucidated by Haldane, extends this phenomenon to scenarios without an external magnetic field, relying instead on the breaking of the time-reversal symmetry {intrinsically}~\cite{HaldanePRL88, QAHRMP23}. 
The modern understanding of this precise quantization is related to the topology of the underlying Hamiltonian. 
For example, the Hall conductivity
$$
\sigma_{xy} = \f{e^2}{h} \mathbb{C}, 
$$
is related to the the Chern number $\mathbb{C}$, a topological index classifying the wavefunction in the momentum space~\cite{TKNNPRL82}. 
The resistance quantization, therefore, is robust against the disorder or shape of the sample.
While we understand the Hall quantization, the theory governing the plateau transitions remains an active research area even now~\cite{HuckesteinRMP95, KramerPR05, EveM08, SlevinOhtsuki12}.
For instance, the Pruisken's $\sigma$-model posits that the fixed points in the two-parameter flow of dimensionless conductivities $\sigma_{xy}$ and $\sigma_{xx}$  describe the IQH critical point~\cite{PruiskenNPB84, PruiskenPRB85, KhmelnitskiiJETP}. 
However, quantitative predictions of the critical parameters within this field theory prove challenging, primarily due to the strong coupling nature of the fixed point $\sigma_{xx}\sim \mathcal{O}(1)$.
As it turns out, perturbative calculations are only feasible in the metallic limit when $\sigma_{xx}\gg 1$.

Therefore, numerical investigations of the critical point are crucial for supporting proposed theories.
The Chalker-Coddington~(CC) network model~\cite{CC_88} stands as the paradigm for numerical simulations and has been instrumental in predicting the localization length exponent $\nuI \simeq 2.55-2.61$~\cite{SlevinOhtsukiPRB09, ObusePRB10,  AmadoPRL11, FulgaPRB11, SlevinOhtsuki12, ObusePRL12, NudingPRB15, DresselhausPRL22}. 
{In tight-binding square lattices -- modeling the lowest Landau level -- this value has been confirmed~\cite{PuschmannPRB2019, Puschmann2021}, while the scaling of the conducting {extended} states in Landau level models predict a slightly smaller exponent $\nuI \simeq 2.4-2.48$~\cite{HuoPRL92, IppotiPRB18, ZhuPRB19}. 
Recently, the IQH transition represented in a dual composite-fermion model on a square lattice was found to be also consistent with CC network calculations~\cite{HuangPRL2021}.
However, predicting the leading irrelevant scaling exponent in numerical simulations remains challenging; nonetheless, multiple studies predict a value $y\approx 0.4-0.6$~\cite{WangPRL96, EversPRB01, MarkosPRL05, ObuseEPL2013}. 
Despite the majority of the work suggesting an exponent within the range of $\nuI \approx 2.5-2.6$, recent conformal field theory studies {have conjectured an} exponent $\nu= \infty$ with $y=0$~\cite{BondensanNP17, Zir19}. 
Such a possibility has been tested on the CC model. Still, with finite-size numerics, it is challenging to conclude the logarithmic scaling of $\sigma_{xx}(L) $ with $L$ towards the fixed point value $\sigma_{xx}^\mathrm{c}=2/\pi$ with certainty~\cite{DresselhausAOP21}.
These factors collectively emphasize the formidable challenge in the numerical investigation of the IQH critical point, primarily due to significant finite-size corrections.

On the contrary, recent experiments observe a critical exponent significantly smaller, $\nu\simeq 2.38$~\cite{KochPRL91, WanliPRL09}.  
One possible explanation for the discrepancy is the presence of electron-electron interactions in experiments. 
Though the short-range interactions are irrelevant at the transition and do not affect the critical properties, the long-range nature of the Coulomb interaction could make a significant change. 
However, such proposals are difficult to verify in numerical studies.
Another possible explanation is the effect of geometric disorder~\cite{GruzbergPRB2017, KlumperPRB19}; such models in the continuum limit could be mapped to Dirac fermions in random potentials or equivalently to a 2D quantum gravity. 
The argument is the following: the regular CC model does not capture all possible disorder potentials that could be present at the IQH critical point; the random network accounts for those missing potentials. 
Indeed, the numerical calculation found an exponent $\nu \simeq 2.37(2)$, which compares well with the experimental observation.

In this work, we revisit this aspect by considering a quantum anomalous Hall transition in an amorphous system devoid of any long-range crystalline order~\cite{AdhipPRL17, MitchellNatPhy2018, YangPRL19, MukatiPRB20, AdhipPRR20, WangPRL21, JuliaPRL22, KimPRL23, GrushinPRL23, ChengPRB23, Corbae_2023, CorbaeExpt23}. 
Our motivation for taking such a model is twofold. 
First, the 2D disordered Dirac fermions model in the continuum limit was conjectured to share the universality class with the IQH transition~\cite{ludwigPRB1994}. 
This connection has been recently investigated numerically in a model of disordered Chern insulators on a square lattice~\cite{sbierskiPRL2021}. 
These investigations reveal that at $E=0$, the localization length exponent is $\nu \approx 2.3-2.36$, similar to the geometrically disordered CC model~\cite{GruzbergPRB2017}. 
However, at finite energies $E>0$, the exponent closely aligns with the anticipated IQH value.

Secondly, we aim to scrutinize whether the QAH transition in the amorphous system deviates from the conventional lattice model. 
% %
This inquiry gets further impetus from recent numerical work indicating that the amorphous topological transition potentially belongs to a different universality class, with the possibility of non-universal exponent values falling within the range $1 \lesssim \nu \lesssim 1.35$~\cite{Ivaki2020PRR} across the topological transition. 
The suggestion is that this unexpected shift in the exponent arises due to the proximity of the percolation transition to the topological transition of the model.  
While the previously studied model belongs to the superconducting class D in the 10-fold classification of topological matter~\cite{RyuNJP2010}, it becomes all the more crucial to investigate such non-universal behavior within the amorphous unitary class A, where presumably, one could expect microscopic details not to affect the critical properties.

While numerous studies have addressed the topological transition in amorphous models~(see Ref.~\cite{Corbae_2023} for a review), a significant gap exists in investigating the critical properties of the QAH transition to the same precision typically performed at the IQH transition using the CC model or tight binding lattices in the presence of an external magnetic field. 
We employ two complementary observables, the two-terminal conductance, and single-particle wavefunction multifractality, to probe the transition and its associated critical properties. 
The disordered average conductance $\meang$  signals the transition from the topological phase with a finite number of edge modes to the trivial phase with $\meang=0$.
At the IQH transition, numerical studies in the CC model found that it is universal $\meang_\mathrm{c} \simeq 0.58(3) \, [e^2/h]$ with considerable mesoscopic fluctuations, $\la (\delta g)^2\ra_\mathrm{c} \simeq 0.081(5) (e^2/h)^2$~\cite{WangPRL96, ChoFisherPRB97, WangLiSoukoulisPRB98, MarkosPRL05}. 
It is interesting to note that the ${\sim} 20\%$ deviation of $\meang_\mathrm{c}$ from the critical transverse conductivity $\sigma_{xy}=1/2\, [e^2/h]$ is not related to numerical errors and is believed to be due to the multifractal nature of the wavefunctions~\cite{MarkosPRL05}.

The different moments of the wavefunction amplitudes are characterized by an infinite set of critical exponents $\tau_q$ defined via the scaling of the inverse participation ratios~(IPR) $\ipr =  \int d{\bs r} |\psi({\bs r})|^{2q} \sim L^{-\tau_q}$, with the system size $L$.
At the IQH transition, the multifractal character of the wavefunction shows up by a non-linear dependence of $\tau_q$ on $q$~\cite{EveM08}.
The critical theories of IQH put severe restrictions on these exponents; for instance, the conformal theory of IQH transition would suggest an exact parabolic spectrum of the anomalous dimensions $\Delta^{\rm IQH}_q \equiv \tau^{(\mathrm{p})}_q - 2(q-1) =  \gamma q(1-q)$, with $\gamma=1/4$~\cite{Zirnbauer1999, Bhaseen00, BondensanNP17, Zir19}.
However, several recent studies have seen substantial deviations from exact parabolicity in exact numerics on the CC model~\cite{EversPRB01,  Evers2008parabolic, Obuse2008}.  
What this would imply for the conformal invariance of the critical point is a matter of an active field of research currently~\cite{zirnbauer1999conformal, BondensanNP17,  Zir19}.

Our study demonstrates that the Chern amorphous system exhibits universal features akin to quantum Hall criticality. 
Specifically, we find an exponent $\nu \simeq 2.60(5)$, consistent with the current precise estimate of $\nu=2.61(1)$~\cite{SlevinOhtsukiPRB09, ObusePRL12, PuschmannPRB2019, DresselhausPRL22}, regardless of the approach taken to the transition at an energy $E=0$ with an irrelevant scaling exponent $y\simeq 0.3(1)$. 
Both multifractal analysis and finite-size scaling of the two-terminal conductance corroborate this estimation. 
Notably, the conductance displays a wide distribution at criticality, reminiscent of class A universality.
Additionally, the multifractal spectrum adheres to the reciprocity relation $\Delta_q=\Delta_{1-q}$ respecting global conformal invariance.
We observe a deviation in $\gamma\approx 0.246(2)$ from its expectation  $\gamma=1/4$; moreover, within the accessible system sizes, the flow towards the asymptotic value with system sizes has a different sign compared to previous data in the CC network~\cite{Obuse2008, Evers2008parabolic}.
Finally, our data supports possible quartic corrections to the parabolic spectrum, with a fixed curvature as the system sizes increase. 
These observations provide a comprehensive view of the critical behavior of the amorphous Chern system.

\section{The Model}

\subsection{Review of different disorder models}
Here, we review different models of disorder that have been used previously to model the IQH critical point and study its critical properties. 

\paragraph*{Continuum potentials:}
 
The continuum description of the disorder, potential $V(\br)$ is usually the starting point of any analytical calculations. It implies modeling of the disorder either with white noise disorder, i.e.,  $\overline{V(\br)V(\br^\prime)} = \delta(\br-\br^\prime)$ or with a finite correlation length such as, 
$\overline{V(\br)V(\br^\prime)} \propto \f{1}{2 \pi \sigma^2} e^{|\br - \br^\prime|^2/2\sigma^2}$, {$\sigma$ is the standard deviation}. 
Similarly, in several numerical works, the lowest Landau level model with continuum potential has been studied, but in many of these works the exponent is slightly smaller than the CC network, see for example, Ref.~\cite{ZhuPRB19}, where $V(\br)=\sum_i \delta(\br -\br_i)$. 

\paragraph*{Lattice models:}
%%%%%%%%%%%%%
The standard model to study the IQH transition is defined on a 2D square lattice, 
$\msr{H} = -t \sum_{\langle ij \rangle} (c^\dagger_i c_j + \mathrm{h.c.}) + \sum_i \epsilon_i c^\dagger_i c_i$, where $\epsilon_i$ represent the onsite disorder, usually taken from uniform distribution. 
In the presence of a magnetic field, the hopping matrix elements are modified via the Peierls substitution.
These crystalline microscopic model of noninteracting electrons usually provides an exponent that is concomitant with the semiclassical network model, for example,  Ref.~\cite{PuschmannPRB2019} or tight-binding model for QAH~\cite{sbierskiPRL2021}.  

 \paragraph*{Network model:}
 The most celebrated model of IQH is the Chalker-Coddington network model~\cite{CC_88}. 
In the limit of sufficiently smooth (on the scale of magnetic length) disorder potential, the electron's motion can be treated within the semi-classical approximations. 
The CC model introduced a network of saddle points connected via the equipotential lines along which the electrons drift and acquire a random phase as they pass through the node.
The model represents the IQH {criticality} for symmetric reflection and transmission amplitudes and has been widely used to study the critical properties; see, for example, Ref.~\cite{EveM08}.
In the continuum limit, the CC model reduces to the nonlinear $\sigma$ model~\cite{zirnbauerJMP1997}.

\subsection{Amorphous lattice}
%%%%%
\begin{figure}[!t]
    \centering
    \includegraphics[width=1\columnwidth]{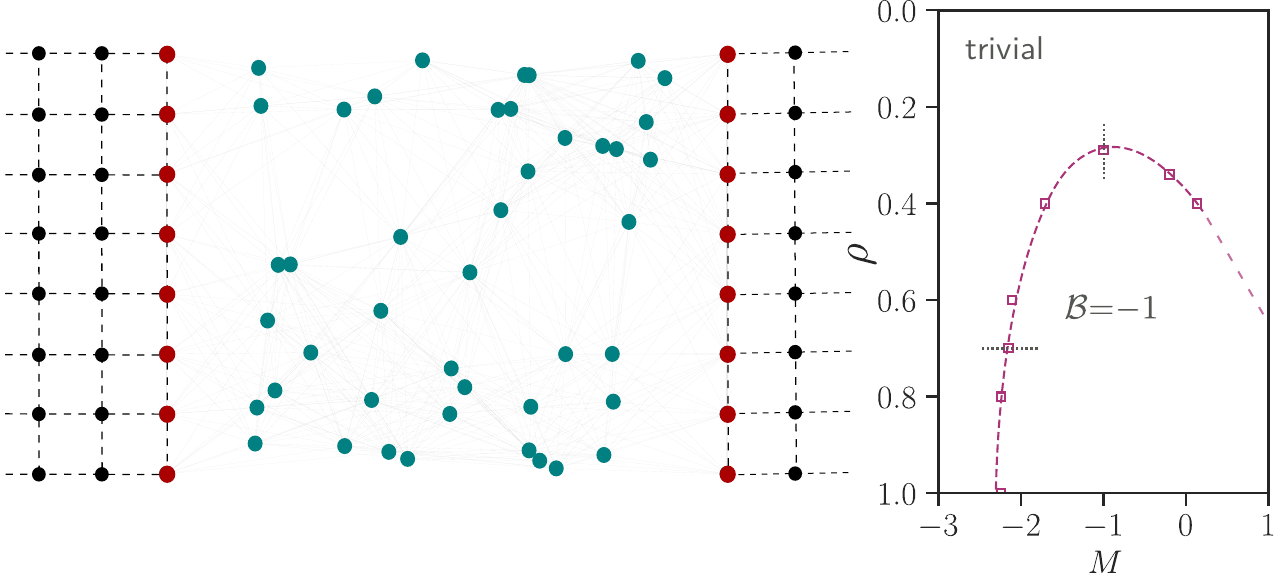}
        \caption{Setup for two terminal conductance  $\meang$ calculations. Square lattice leads are connected to a scattering region with random lattice points. The system shown here has system size $L=8$ with a density of lattice points $\rho=0.7$ and with open boundary condition perpendicular to the transport direction. The gray lines connect the lattice sites within a hopping distance of $R=4$. One extra slice of regular lattice sites (indicated with red dots) is embedded into the scattering region on both sides for a smooth connection to the ideal square leads.
        Right panel: Shows the qualitative topological phase diagram of the model~\eqref{eq:ham}. The symbol is calculated using the Bott index (for system size $L=48$ and a single lattice configuration), and the dotted line is a guide to the eye. 
        } 
    \label{fig:geometry}
\end{figure}
On the contrary, an amorphous lattice has no crystalline order, as is the case for a regular tight-binding model. 
The amorphous models are constructed by placing random $N$ points confined to a 2D square box of area $L^2$ characterized by the density $\rho = N/L^2$. 
The real space coordinates of the lattice points are random variables sampled from a uniform distribution $x,y\in [0, L]$. Each site accommodates 2-orbital degrees of freedom. 
Fermions hop on this random graph to sites within a circular ring of radius $R$. The model was introduced in Ref.~\cite{AdhipPRL17} and has the following form, 
\eq{
\msr{H} = - \sum_{ i,\alpha;j,\beta } T_{ \alpha \beta} ({\bm{r}_{ij}})c_{i,\alpha}^{\dagger}c_{j,\beta },  
\label{eq:ham}  
}
with the orbital hopping matrix given by,  
\eq{
 T_{\alpha\beta} (\br_{ij}) = \Vec{d} \cdot \Vec{\sigma}\, \, \mathrm{C} e^{-\br}\Theta({\bm R}-\br),  % not \bm here as r and R are magnitude only 
\label{orbita_structure}
 }
 with, 
\eq{
    d_0 &= \f{t_0}{2}, \quad \quad d_x = -\frac{1}{2} \cos{\theta} (i + \cos{\theta})\nn \\
    d_y &= \frac{i}{2}\sin{\theta}( \frac{i}{2}\sin{\theta}-1 ), \quad \quad   d_z =  -1/2,  \nn 
}
for $\bm{r}_{ij}\ne0$ (inter-orbital hopping) and $(d_0,d_x,d_y,d_z)=(0,1/2,1/2,2+M)$ for $\bm{r}_{ij}= 0$ (intra-orbital hopping). Here $\Vec{\sigma}=(\sigma_0,\sigma_1,\sigma_2,\sigma_3)$ and $\Vec{d}  = (d_0,d_x,d_y,d_z)$ are 4-component vectors with $\sigma_0 = \mathds{1} $ and $\sigma_i$ is the $i^{th}$ Pauli matrix. And, $\bm{r}_{i}$ is the position vector of the $i^{th}$ lattice site (w.r.t a fixed Cartesian system), $\bm{r}_{ij}=\bm{r}_i-\bm{r}_j, |\br_{ij}| = r$, $\theta$ being the angle that the vector $\bm{r}_{ij}$ makes with the positive x-axis. $\mathrm{C}=e~ \text{for}~ \bm{r}_{ij}\ne 0$, else $\mathrm{C}= 1$. The orbital hopping matrix is modulated by an exponential decay with distance away from a lattice site. $M$ is the mass parameter that creates the difference in onsite energy and breaks the sublattice symmetry.
%the parameter $\lambda$ tunes the intra-orbital coupling.
The Hamiltonian, not being a real symmetric matrix, breaks the time-reversal symmetry. A finite $t_0 \ne 0$ breaks the charge-conjugation symmetry and renders the system in unitary class A. The parameter $t_0$ distinguishes model~\eqref{eq:ham} from that studied in~\cite{Ivaki2020PRR}.  For the simulations done in this work, we keep  $t_0 = 1/4, R=4$ fixed and vary the parameters $M$ and $\rho$ to probe the scaling behavior.

The system hosts a non-trivial topological phase characterized by the real-space invariant, Bott-index~\cite{Loring_2010}, in the parameter space of the Hamiltonian as shown in Fig.~\ref{fig:geometry} right panel.
We study the critical properties of the topological transition in two different point of the phase diagram (marked with dashed lines). 
For the conductance calculation, the lead is connected to the left and right edges of the system, as shown in Fig.~\ref{fig:geometry}. 
% \NP{boundary condition}
The periodic boundary is used for the multifractal calculation at the critical point. 

\section{Observables}
\subsection{Conductance}
In two-terminal conductance measurements, infinite translation-invariant leads are connected to a finite phase-coherent scattering region, the latter modeled by tight-binding Hamiltonians of the form in Eq~(\ref{eq:ham}). The electron wavefunctions in the leads are plane waves $\phi_n = \chi_n e^{\pm ik_n} ,$ where $k_n$ is the longitudinal momentum for the $n^{th}$ propagating mode, $n = 1,2,3, \ldots, N$ and $\chi_n$ is the transverse component of the wavefunction. The $2N$ propagating modes (for left and right lead), also called scattering channels form  a basis for the incoming waves $(\psi^{\mathrm{in}})$ and the scattering matrix $S$ transforms this incident wave in the left lead to outgoing waves $(\psi^{\mathrm{out}})$ in the right lead as follows, 
\begin{equation}
   \psi^{\mathrm{out}}  = S  \psi^{\mathrm{in}},\, S =  \begin{pmatrix} 
	r & t'\\
	t & r' \\
	\end{pmatrix}, 
\end{equation}
with $N\times N$ reflection matrices $r$ and $r'$ (from
left and right boundary) and transmission matrices $t$ and $t'$ ( from left to right and from right to left). The transmission matrix $t$ allows the calculation of differential conductance by Landauer formalism and is  expressed as 
\begin{equation}
   g = \frac{e^2}{h} \mathrm{Tr} (t^{\dagger}t),
   \label{eq:cond} 
\end{equation}
where $g$ is the dimensionless conductance (in units of $e^2/h$).

The transition to an IQH state is characterized by the appearance of quantized conductance plateaus. 
We adopted the longitudinal conductance $\meang$ (overline denotes the disorder averaging) as an order parameter to study the zero-temperature quantum phase transition that occurs as we traverse between phases of different topological numbers. 
The conductance for the model studied is a function of model parameters, density, and energy, i.e., $\meang(L; M,\rho,E)$. 
The parameters $M,\rho,E$ form a critical surface in the parameter space, where the system is IQH critical~\cite{Moreno2023}.
We study only the $E=0$ critical conductance by tuning the parameters $M$ and $\rho$, and the energy dependence of the $\meang$ at the critical point is presented in the appendix. 
 
We utilize a square geometry of size $L\times L$ (with the largest system size being $L=768$) as the scattering region for conductance simulations at $E=0$. 
The two-terminal setup is shown in Fig.~\ref{fig:geometry}. 
To maintain a smooth connection with the lead, we added an extra layer of regular sites at the end of the sample, maintaining the density of sites in the y-direction fixed. 
For a given choice of parameters, conductance is calculated using the open-source quantum transport package {\tt Kwant}~\cite{Groth2014Kwant}. 

\subsubsection{Finite size scaling}

Finite-size scaling of observables at an integer quantum Hall critical point poses serious numerical challenges. 
Several numerical works underestimated the localization length exponent $\xi \sim |x|^{-\nu}$ by assuming a one-parameter scaling collapse $F(L^{1/\nu}x)$, for some tuning parameter $x$ ignoring the irrelevant correction to scaling.
Since the work of \textcite{SlevinOhtsukiPRB09}, the importance of irrelevant scaling operator at IQH critical point is well understood; the correction is more significant for smaller system sizes.

The order parameter $\meang$ (or equivalently $\overline{\ln g})$ is expanded in terms of leading relevant scaling observable to an expansion order $N_R$ given by
\eq{
\meang = \msr{F}_0(x) + b_0 L^{-y} \, \msr{F}_1(x) + c_0 L^{-2y} \, \msr{F}_2(x) 
\label{eq:g_expand}
} 
where, 
$
\msr{F}_j = \sum\limits_{n=0}^{N_\text{R}} a_{j n} x^n
$
, and $x = (M-M_\text{c})/M_\text{c} \cdot  L^{1/\nu}$, $M_\text{c}$ being the critical mass parameter, $\nu$ and $y$ are the leading relevant and irrelevant exponents respectively and $(b_0,c_0,a_{jn})$ are expansion coefficients. 
The irrelevant expansion is kept to a maximum order of two to limit the number of fitting parameters; it turns out that at least two irrelevant scaling variables are needed to address the correction to scaling. 
The expansion has $3(N_\text{R}+1)+3$  number of unique fitting parameters ($b_0, c_0,a_{jn},\nu, y, M_\text{c}$).
We studied the stability of the fit by varying the order of expansion $N_\text{R}$ and also the order of the irrelevant expansion in our analysis, see App~\ref{app:fit} for further details. 
The fitting procedure gives the best estimates of fitting parameters, and Eq.~\eqref{eq:g_expand} can be used to reconstruct the scaling function $\meang$.

To observe the scaling collapse, the irrelevant contribution to the order parameter is subtracted to get the corrected order parameter ($\gcorr$) in terms of the relevant scaling operator as, 
\eq{
\gcorr = \meang - b_0 L^{-y} \, \msr{F}_1(x) - c_0 L^{-2y} \, \msr{F}_2(x). 
\label{eq:gcorr_expand}
}
When plotted with the parameter $\tilde{M}L^{1/\nu}$, this function shows a one-parameter scaling collapse, and the fitted function comes as a byproduct of this scaling analysis.

\subsection{Multifractality}

The multifractal spectrum with the exponents $\tau_q$ is a characteristic fingerprint of the critical point of Anderson transitions~\cite{EveM08}.
It describes the system-size scaling of  moments of wavefunctions $\psi({\br})$ at criticality:  
\eq{
\overline{P}_q = \sum_{{\br}}\left(\sum_{\sigma=\uparrow,\downarrow}\mid \Psi_\sigma(\br)\mid^{2}\right)^q \sim L^{-\tau_q}; 
%\label{boxweight}
\label{momentscaling}
} 
The overline indicates an average over an ensemble of disorder samples and $\sigma$ the two degrees of freedom at each lattice site.

In contrast to previous works, where the multifractal spectrum usually characterizes the critical point, here, we generalize the notion of the critical MF exponents $\tau_q$ to an $\textit{effective}$ exponent, even away from the critical point, where $\tau_q\to 0$ in the limit of large system sizes $L$, indicating Anderson localization and hence the absence of scaling with of the moments with $L$.

Similar to works about conventional localization-delocalization transitions \cite{RodVSR11,EveM08} and topological phase transitions in other symmetry classes \cite{PusHLB21}, we employ a system-size scaling approach. 
We define effective multifractal exponents as the logarithmic finite difference between subsequent system sizes,
\begin{align}
	\tilde{\tau}_q(X,L)&=\frac{\ln \langle  
    P_q(X,L)
    \rangle -\ln \langle P_q(X,L/2)\rangle}{\ln L- \ln{L/2}}~,\label{eq:mfa_tau}
\end{align}
where $X$ is the tuning parameter across the phase transition in the vicinity of the critical point at $X_{\mathrm{c}}$~\cite{RodVSR11,Moreno2023,Altland2023,Puschmann23}.  
Here, $\tilde{\cdot}$ denotes effective multifractal exponents: It is defined even for nonfractal states away from criticality, and they are obtained for a finite system of size $L$. In that sense, they are not actual exponents governing a power law scaling. 
At the critical point for large $L$, the  $\tilde{\tau_q}$ reduces to an actual exponent as in Eq.~\eqref{momentscaling}; and $\tau_q=\lim_{L\to\infty} \tilde{\tau}_q(L)$ yields the multifractal spectrum of the IQH transition. 

To extract the scaling of the localization length close to the critical point, we monitor the system size dependence of the \textit{effective} MF exponent $\tilde{\tau}_q$. It should be stable at the critical point and diverge towards zero in the phases nearby. The asymptotics of the functional dependence of the divergence towards zero, depending on the distance to the critical point, will allow us to extract the localization length exponent $\nu$. 

The (effective) anomalous dimension is defined as
$\tilde{\Delta}_q=\tilde{\tau}_q-d(q-1)$. 
In particular, it has been hypothesized that at the critical point, the anomalous MF dimension should be exactly parabolic with
\begin{equation}
    \Delta_q^{\rm IQH}=\gamma\, q (1-q) \text{   with   } \gamma=\frac{1}{4},
    \label{eq:deltaq}
\end{equation}
which serves as our benchmark to characterize the QH critical point~\cite{Zir19}.

Numerically, the linear system size $L$ of the Hamiltonian~\eqref{eq:ham} is scaled between $16$ and $768$. We use the implicitly restarted Lanczos method with shift invert at zero energy - the routine {\tt ARPACK} as implemented in scipy~\cite{ARPACK} - to calculate the five lowest lying eigenvectors of the finite size Hamiltonian matrix. The number of disorder realizations considered per system size is summarized in Table~\ref{tab:dis_config}.

\section{Numerical Results}

\subsection{Conductance}
\begin{figure}[!bt]
    \centering
    \includegraphics[width=1\columnwidth]{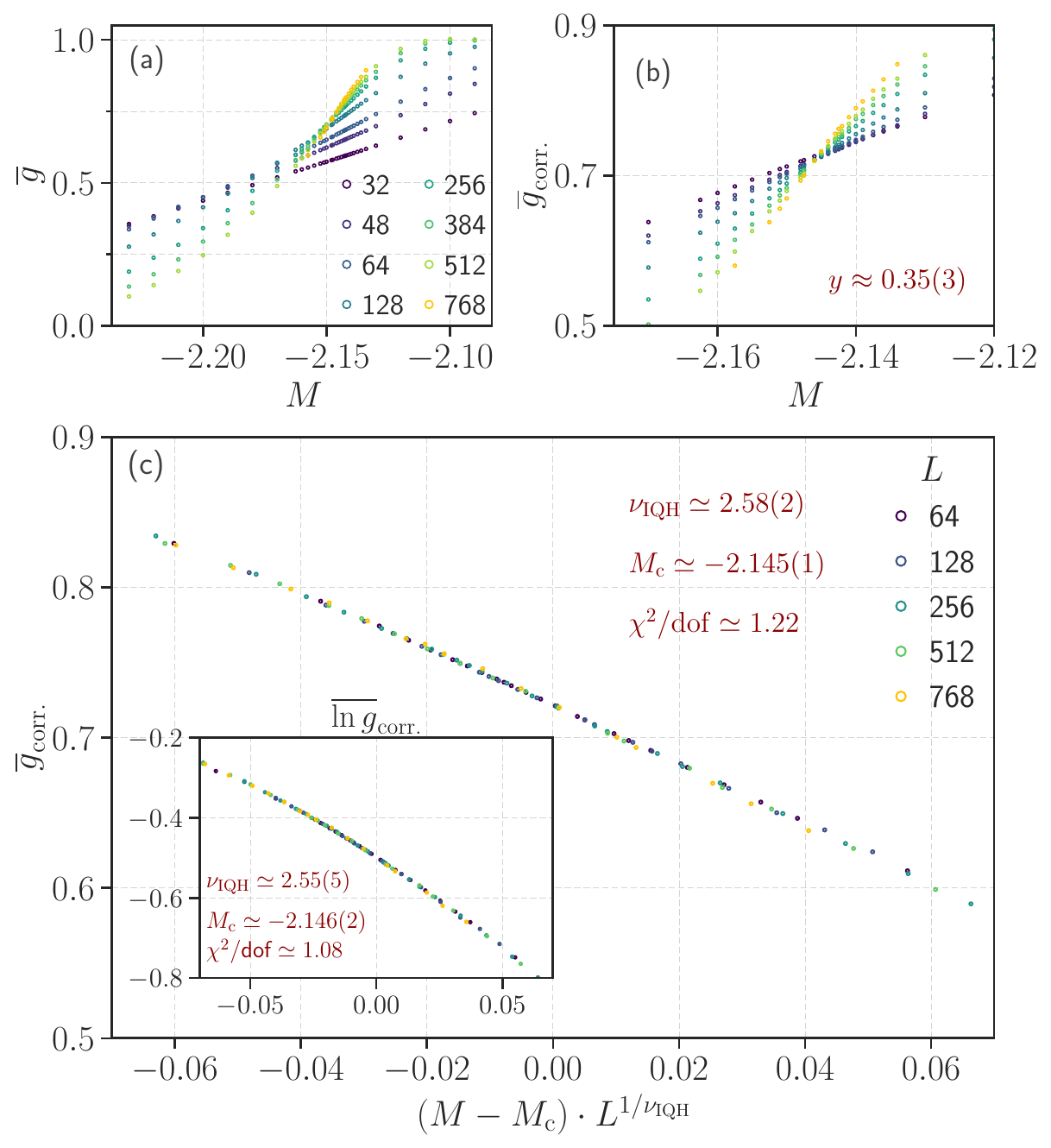}
    \caption{Finite size analysis of the mean conductance $\meang$ and $\overline{\ln g}$. (a)~Show the raw conductance data as a function of the mass parameter $M$ for different system sizes $L=\{32-768\}$ with density $\rho=0.7$. Shift in the crossing point with successive system sizes indicates a strong finite size effect. (b)~The corrected conductance $\gcorr$ using Eq.~\eqref{eq:gcorr_expand} with $N_\mathrm{R}=2$. (c)~Show the scaling collapse of the data with $\nu\simeq 2.58(2)$, and $\Mc\simeq-2.145(1)$ with a $\chi^2\approx 1.22$. For better visibility, only a few system sizes are shown. Inset shows the scaling collapse of the $\overline{\ln g}_\mathrm{corr.}$ with similar leading exponent $\nu\approx 2.55(5)$. The data collapse is obtained with an irrelevant scaling exponent $y\approx 0.36$. }
    \label{fig:g_collapse}
\end{figure}

The raw data for mean conductance $\meang$ close to the transition, scaling with system size $L$ and mass parameter $M$ is shown in Fig.~\ref{fig:g_collapse}(a). The data close to the critical point has relevant and irrelevant contributions, with the irrelevant scaling being more pronounced for the smaller system sizes. This is evident from data not showing a unique crossing point, implying that the asymptotic limit has yet to be reached. 
We extract the critical parameters $(\nu, y, M_\text{c}, a_n, b_n)$ by fitting the data to the scaling form in Eq.~\ref{eq:g_expand} through a $\chi^2$-minimization procedure. %%
The fitting parameters thus obtained, along with the $\chi^2$ value, are shown in table~\ref{tab:cond}. We estimate the leading relevant exponent for this transition to be $\nu \simeq 2.60(5)$, which agrees reasonably well with other numerical studies of integer quantum Hall criticality. The fitted value of the irrelevant scaling exponent is $y\simeq 0.3(1)$.

With the estimated scaling parameters from the above fitting procedure, we estimate the pure relevant scaling near the critical point using Eq.~\ref{eq:gcorr_expand}. This corrected conductance $\meang_\mathrm{corr.}$ data is presented in Fig.~\ref{fig:g_collapse}(b). The data here shows only the dependence on relevant scaling variables and offers a unique crossing point as a function of $L$ and $M$. This corrected data shows a one-parameter scaling collapse with the joint variable of $M$ and $L$. This is shown in the lower panel of Fig.~\ref{fig:g_collapse}(c) along with the scaling function, and we get an excellent collapse with a $\chi^2 \approx 1.2$. 

The scaling collapse for the observable $\overline{\ln g}$ is displayed in the inset in Fig.~\ref{fig:g_collapse}(c), along with the estimated critical parameter is $\nu\simeq 2.55(5)$, which is, however, slightly smaller than the $\meang$ scaling but overlaps within our 1 $\sigma$ errorbars. 
We believe the discrepancy comes from the instability associated with the multi-parameter fits as documented in Tab.~\ref{tab:cond}, \ref{tab:cond2}.

\begin{figure}[t]
    \centering
    \includegraphics[width=1\linewidth]{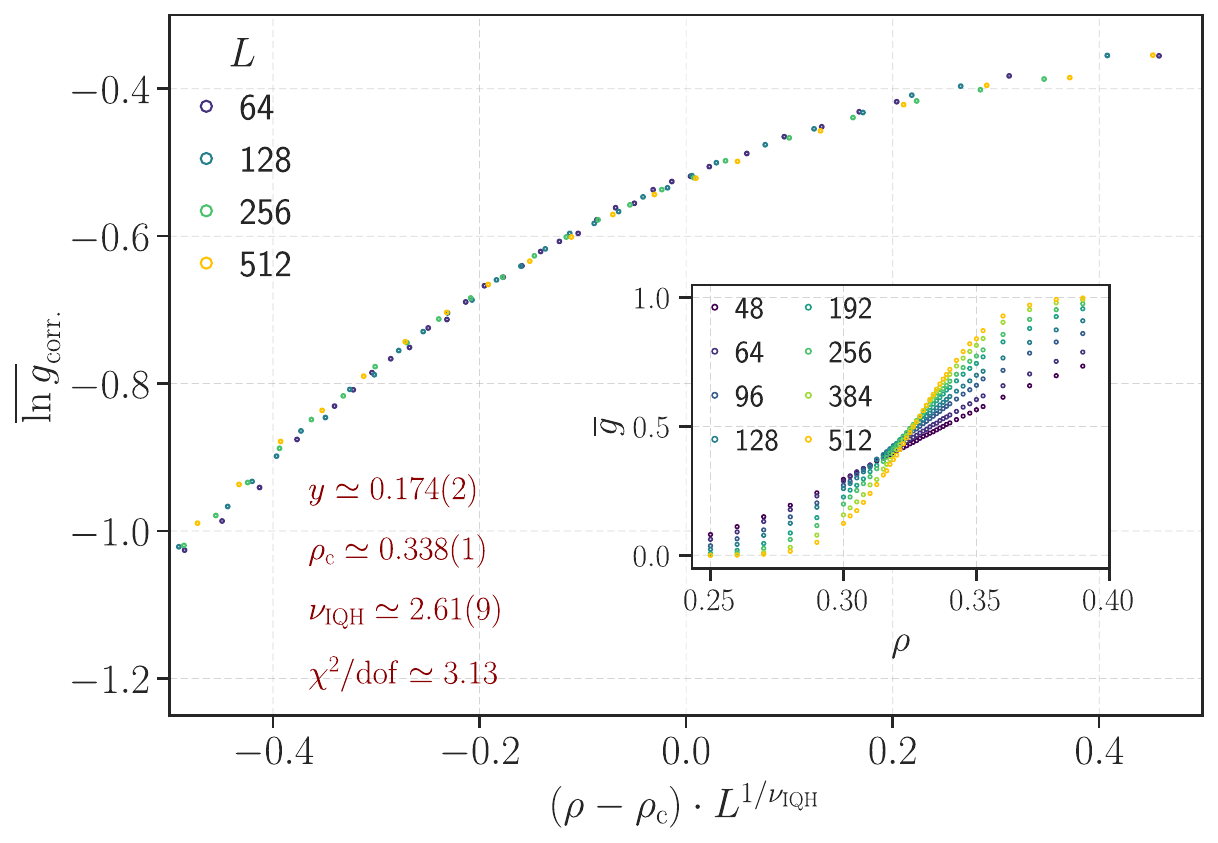}
    \caption{Finite-size scaling analysis of the log-conductance $\overline{\ln g}$ by varying the density $\rho$. The mass parameter is fixed at $M=-1$. The main panel shows the scaling collapse of $\overline{\ln g}_\mathrm{corr.}$ after subtracting the irrelevant correction (similar analysis as in Fig.~(\ref{fig:g_collapse})). The critical parameters obtained are shown in the figure and table~\ref{tab:cond}. 
    The inset figure shows the raw data at various densities and the system sizes $L=\{64-512\}$.}
    \label{fig:g_collapse_with_rho}
\end{figure}
\subsubsection{Varying the density}
%%%%%%%%%%%%%
Additionally, we monitor the topological transition at a different point on the critical surface by changing the density of particles $\rho$ in the system, keeping the area $A$ fixed. We keep the mass parameter fixed at $M=-1$. The data for this simulation is presented in Fig.~\ref{fig:g_collapse_with_rho}. The scaling analysis follows a procedure similar to that in Fig.~\ref{fig:g_collapse}. The results of the scaling analysis are presented in table~\ref{tab:cond2}. The estimated irrelevant exponent is  $y\approx 0.174(2)$, and the leading relevant exponent agrees with the analysis in Fig~\ref{fig:g_collapse}.  
We found a similar  $\nuI \approx 2.61(9)$ for this transition; however, within the error bars, it is not different from the exponent found with varying $M$. 
However, the $\chi^2$ is significantly bigger for this transition due to the unavailability of larger $L$ at this transition. 
%
%%%%%%%%%%%%%%%%%5

%%%%%%%%%%%%%%%%%%%%%%%%%%%%%%%%%%%%%%%%%5
\begin{figure}[!t]
    \centering
    \includegraphics[width=1\columnwidth]{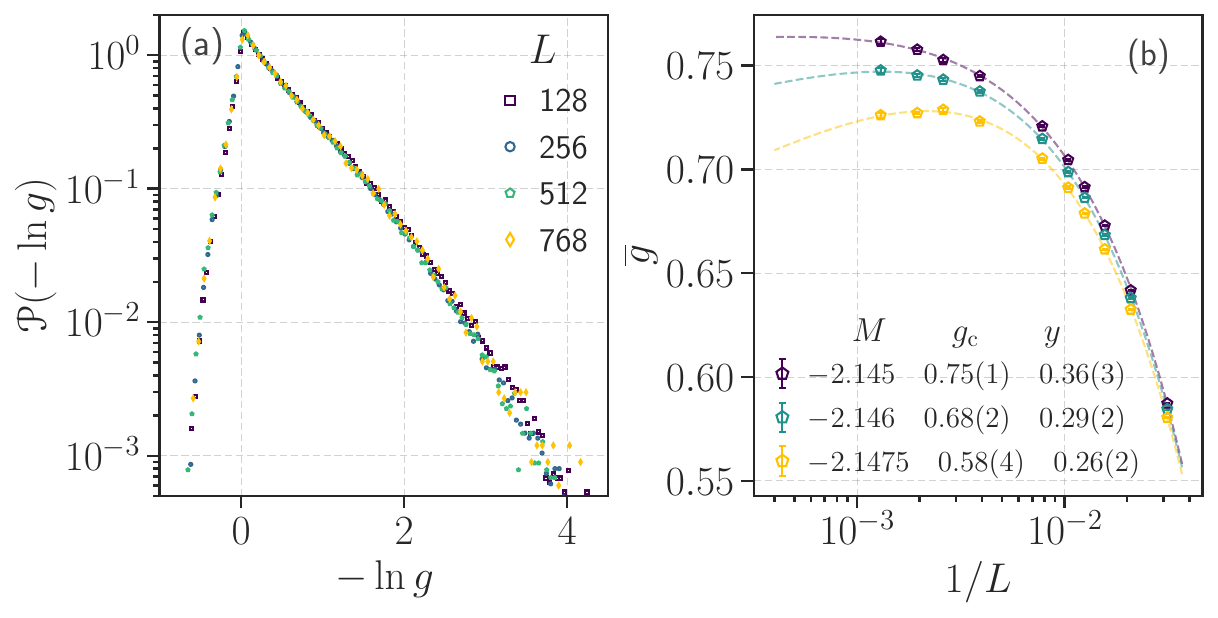}
    \caption{(a)~Shows the scale-invariant conductance distribution at the estimated critical point $\Mc\simeq -2.146$. For small system sizes, the finite size effects are visible in the tail of the distribution. (b)~Highlighting the scaling of the $\meang$ close to the critical point. The solid line indicates the two correction terms mentioned in the legend with $y$ given in the legend. Depending on the precise value of the $\Mc$ the $y$ also changes as the $g_\text{c}$.}
    \label{fig:dist_g}
\end{figure}
%%%%%%%%%%%%%%%%%%%%%%%%%%%%%%%%%%%%%%%%%

%%%%%%%%%%%%%%%%%%%%%%%%%%%%%%%%%%%%%%%%%5
\subsubsection{Conductance distribution}
The fluctuation at the critical point is monitored, and the data for log-conductance ($\ln g $) distribution is presented in Fig~\ref{fig:dist_g}(a). The IQH critical point is marked by a scale-invariant conductance distribution with a wide range of conductance values showing strong fluctuation. 
The estimated first and second moments of this distribution are found to be $\meang\simeq 0.67(2) [e^2/h]$ and $\text{var}(g)\simeq 0.08 [(e^2/h)^2]$. 
{The distribution peaked around $\sim e^2/h$ indicating that with the open boundary, one transport channel (the edge mode in the Chern model~\eqref{eq:ham}) is dominating the transmission; this is reminiscent of the conductance distribution $P(\ln g)$ in quasi-one-dimensional geometry~\cite{MuttalibPRL99}.}

The data shown in Fig~\ref{fig:dist_g}(b) observes the correction to these moments with the irrelevant scaling exponent $y$ has strong variation depending on the variability of $\Mc$ that arises. 
Nonetheless, the observed $y \lesssim 0.4$ is smaller. 
It is important to note that the mean conductance data shows corrections with two irrelevant terms with opposite signs within the system sizes studied. 
The reported variance at the critical point matches reasonably well with the reported values in the literature~\cite{WangPRL96}, and is consistent with the fact that the variance is almost an order of magnitude smaller than the mean conductance.

%%%%%%%%%%%%%%%%%%%%%%%%%%%%%%%%%%%%%%%%%5
\subsection{Multifractality}
%%%%%%%%%%%%%%%%%%%%%%%%%%%%%%%%%%%%%%%%%5
\begin{figure}[!tb]
    \centering
    \includegraphics[width=1\columnwidth]{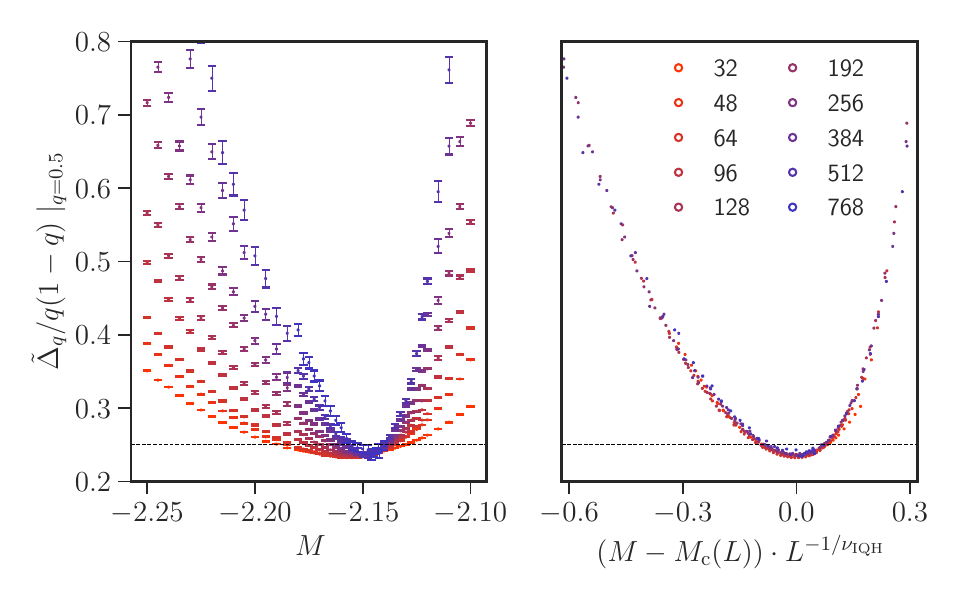}
    \caption{Effective anomalous MF exponent across the IQH transition at $\rho=0.7$ for the moment $q=0.5$ for several system sizes (left) and approximate collapsed with $\nu_\mathrm{IQH}\sim 2.6$ (right). The horizontal line indicates the value of $\Delta_q/q(1-q) \rvert_{q=0.5}$ for parabolic prediction at the IQH transition~\eqref{eq:deltaq}. The finite size correction is taken into account with the $L$ dependent critical mass term $\tilde{M}_\mathrm{c}(L)$. 
    }
    \label{fig:nf_collapse}
\end{figure}

\subsubsection{Localization length exponent $\nu$}
In Fig.~\ref{fig:nf_collapse}, we show the effective MF exponent at $q=0.5$ across the QH transition at $\rho=0.7$. Far from the transition at $\Mc\approx -2.144$, the effective exponent does not converge to a finite value when increasing the system size. This indicates, as expected, Anderson localization inside the trivial ($M > \Mc$) and the topological phase ($M< \Mc$). However, at criticality, the effective exponent for different system sizes coincides (up to irrelevant corrections) approximately at the QH critical value, i.e. $\Delta^\mathrm{QH}_q \sim\frac{1}{4} \, q(1-q)$, also found in the transport calculation.

Here, we analyze the vicinity of the critical point. In Fig.~\ref{fig:nf_collapse} (right panel) we rescale the tuning parameter $M \to (M-M_\mathrm{c}(L))\cdot L^{1/\nu}$ and observe a collapse with $\nu_{\mathrm{IQH}}\sim 2.6$. As the effective dimension $\tilde{\tau}_q$ is a direct measure of localization, this indicates the localization length exponent to be consistent with the results from the conductance calculation and with universality across different models of the IQH effect. 

We extract the curvature of the data shown in Fig. \ref{fig:nf_collapse} to visualize residual irrelevant finite size corrections and analyze its finite size scaling. To this end, we fit the data with a polynomial function of $3$rd order.

\begin{align}
    \tilde{\tau}_{q} &= \tau_q + \tilde{\tau}_q^{\prime\prime} (M-M_\mathrm{c})^2 + \tilde{\tau}_q^{\prime \prime \prime} (M-M_\mathrm{c})^3 
    \label{e4}
\end{align}
with four fitting parameters, $\tau_q, \tilde{\tau}_q^{\prime\prime}, \tilde{\tau}_q^{\prime \prime \prime}$ and $M_\mathrm{c}$. 
Asymptotically, the curvature should scale with the localization length exponent, $\tilde{\tau}_q^{\prime\prime}\sim L^{2/\nu}$.
For finite sizes, we assume additional irrelevant scaling corrections with at least one exponent $y$ of the form 
\begin{equation}
    \tilde{\tau}_q^{\prime\prime} \sim L^{2/\nu} \cdot \Le( 1+ a \, L^{-y} + b \, L^{-2y} \ldots \Ri).
\end{equation}
In Fig.~\ref{fig:loclength} the fitted curvature $\tilde{\tau}_q^{\prime\prime} $ for different $q=0.5, 1.5$ is shown, where the asymptotic finite size scaling, assuming $\nu_\mathrm{IQH}$, has been subtracted. 

Ideally, if universality were true and irrelevant corrections were excluded, the data should follow a horizontal line. In reality, irrelevant corrections are present. However, the data is consistent with the assumed exponent $\nu_\mathrm{IQH}\sim 2.6$; additional finite size corrections can explain the deviations. The irrelevant corrections can be fitted with a relatively large inaccuracy with an exponent $y\approx0.6(2)$ 
%%%%%%%%%%%%%%%%%%%%%%%%%%%%%%
\begin{figure}[!tb]
    \centering
    \includegraphics[width=1.0\columnwidth]{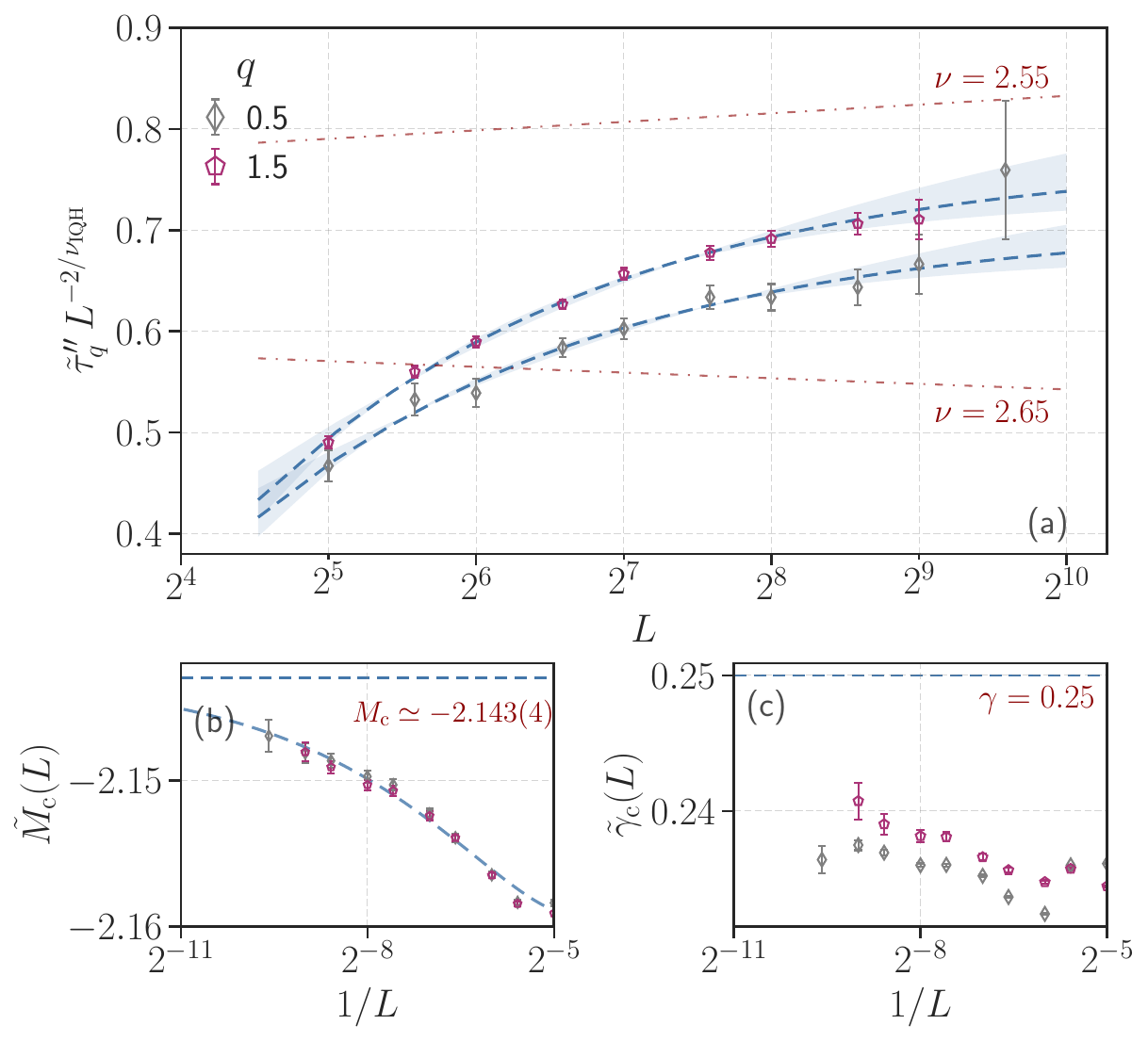}
    \caption{(a)~Shows the residual finite size dependence of the renormalized curvature $\tau^{\prime \prime}_q \cdot L^{-2/\nuI}$ as a function of system size $L$ for two different values of $q={0.5, 1.5}$ with $\nuI \approx 2.6$. 
    The shaded region gives a variability of $y$ between $ 0.4 \lesssim y \lesssim 0.8$. The dashed line indicates a $y=0.6$. 
    (b)~The $1/L$ dependence of the finite size critical mass parameter $\tilde{M}_\mathrm{c}(L)$. The dashed line indicates the $\Mc\simeq -2.143(1)$ obtained from the interpolation of the data fitting two subleading corrections with $y\approx 0.6$. (c)~The finite size dependence of the prefactor $\tilde{\gamma}_q(L)$ of the quadratic term of $\Delta_q$ as defined in Eq.~\ref{eq:deltaq}.
    }
    \label{fig:loclength}
\end{figure}
%%%%%%%%%%%%%%%%%%%%%%%%%%%%%%

In the lower panels of Fig.~\ref{fig:loclength} show the convergence of the remaining relevant fitting parameters, i.e., the critical point $M_\mathrm{c}$, and the MF dimension itself is shown. 
Seemingly, the latter lies consistently below the expected value, assuming parabolic multifractality (dashed line), and also significantly below the value observed in CC network simulations. \cite{Obuse2008,Evers2008parabolic}As this raises questions about the universality of the CC result for this amorphous system, we discuss this in the following by analyzing the MF dimensions at the critical point in closer detail.

\subsection{Multifractal properties at the critical point}
\subsubsection{Reciprocity}
%%%
%%
%
It was shown that at the Anderson critical point, the symmetry relation between multifractal exponents generically is exact~\cite{MirlinPRL06}. This was checked numerically at the IQH transition previously~\cite{Evers2008parabolic}. 
The symmetry $\Delta_q = \Delta_{1-q}$ implies that the following ratio, 
\eq{
\overline{r}_q(L) = L^{2 \, (2q -1)} \cdot \f{P_q}{P_{1-q}} 
}
will cancel the leading $L$ dependence. 
Further, subleading corrections are expected to be small in the asymptotic limit.  
Thus, it is a way to diagnose whether the numerically accessible system sizes are large enough to be in the scaling regime. 
However, recent works suggest that reciprocity develops before the true asymptotic limit of QH criticality is achieved~\cite{GruzbergPRB13}; therefore, solely depending on the convergence of the reciprocity relation could be misleading and has to be taken with care.
Nonetheless, Fig.~\ref{fig:reciprocity} demonstrates the reciprocity relation for different $q$ values. 
With increasing system sizes $L={32-768}$, the $\overline{r}_q$ becomes almost independent of the system sizes within residual statistical noise. 
%
%%%%%%%
\begin{figure}[!t]
    \centering
    \includegraphics[width=\columnwidth]{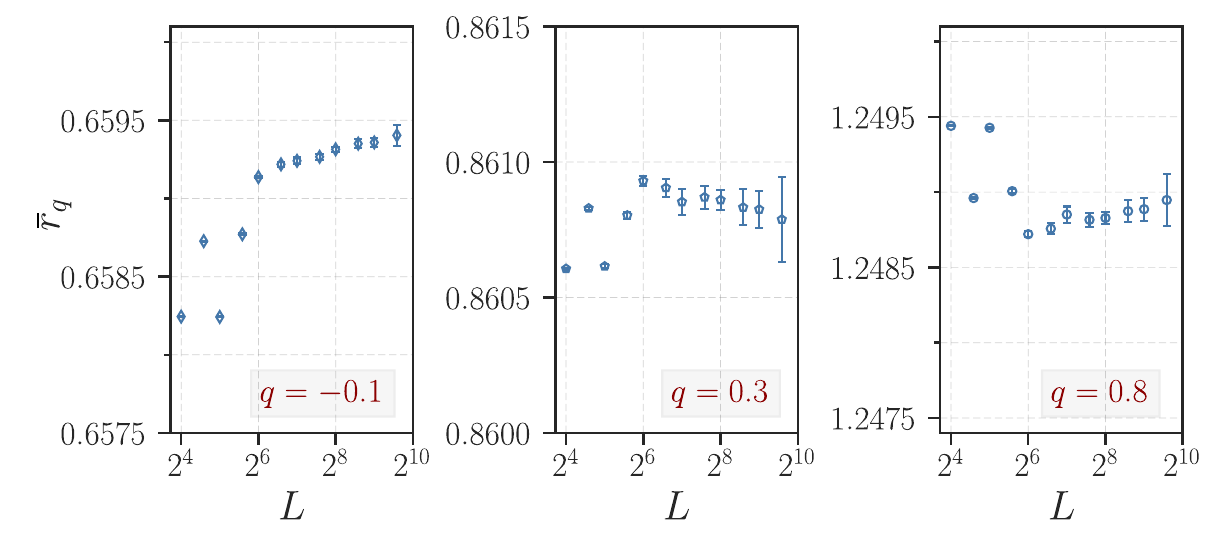}
    \caption{Verifying the reciprocity relation $\overline{r}_q$ at the critical point for $q={-0.1, 0.3, 0.8}$. The convergence is seen in the already for $L \gtrsim 2^7$, with slight deviations, less than $0.1$\%, possibly indicating residual statistical noise.    }
    \label{fig:reciprocity}
\end{figure}

\begin{figure*}[!tb]
    \centering
    \includegraphics[width=1\linewidth]{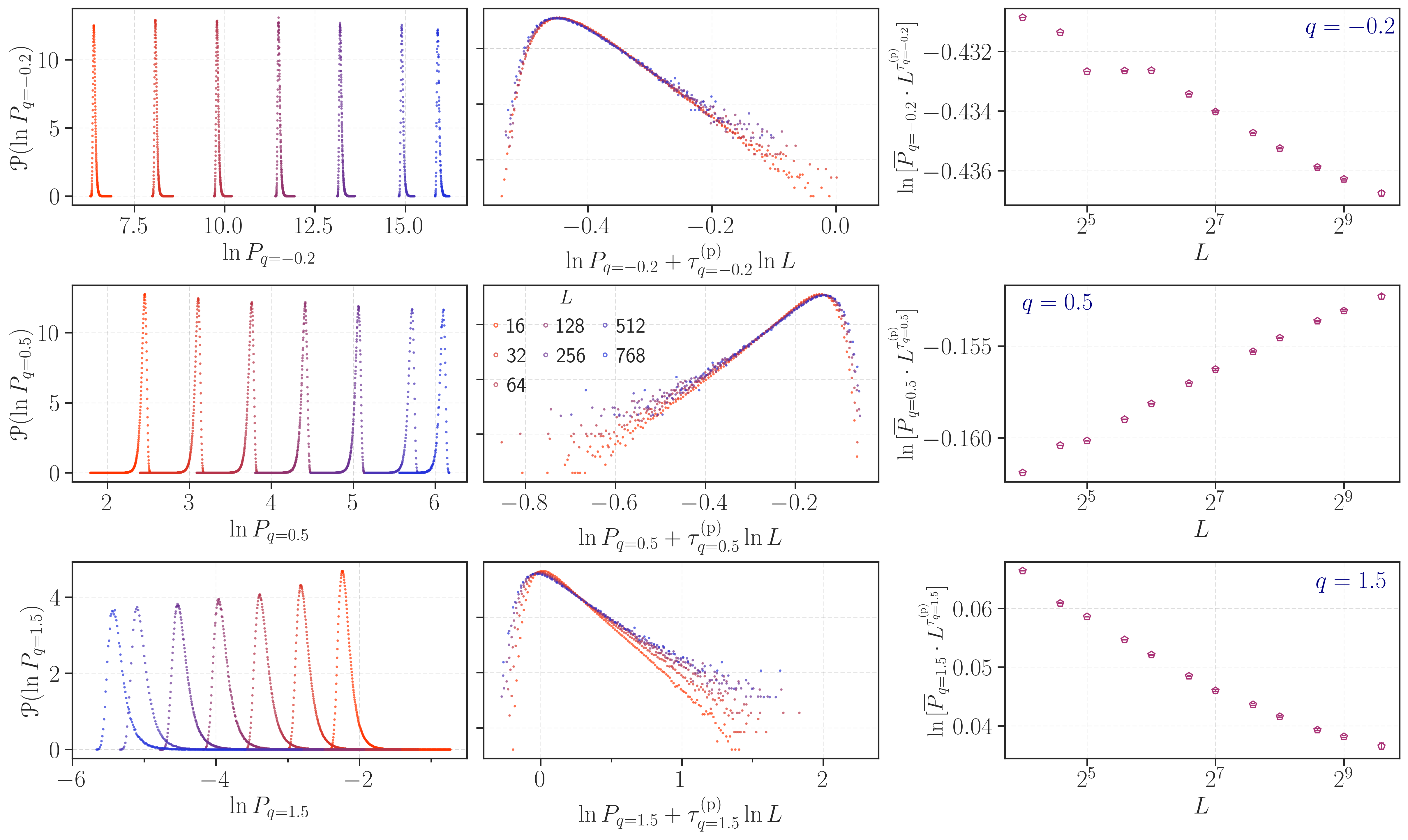}
    \caption{Showing the flow of the distribution of $P_q$ for three different $q={-0.2, 0.5, 1.5}$ with system sizes at the critical point $\Mc \simeq -2.144$ for different system sizes $L=\{16-768\}$. First column: Shows the raw IPR data with system sizes. Second column: Shows the distribution of the logarithm of reduced IPR,  $\ln (P_q \cdot L^{\tau^{(\mathrm{p})}_q})$, assuming a parabolic multifractal spectrum~\eqref{eq:deltaq}. The third column shows the residual system size dependence of the reduced IPR with system sizes. The reduced IPR is expected to saturate in the asymptotic limit for all $q$ values for a parabolic spectrum. We observe significant finite-size corrections for all $q$-values.}
    \label{fig:mfa_distros}
\end{figure*}

\subsubsection{Probing parabolicity: IPR}

We compare the wave function statistics to the prominent parabolic prediction~\cite{Zirnbauer1999, Bhaseen00, BondensanNP17}. The IQH criticality was predicted to come with an exactly parabolic multifractal spectrum with $\gamma=1/4$. 
This, in turn, was questioned by several numerical studies of the IQH transition, all in Chalker-Coddington networks \cite{Evers2008parabolic, Obuse2008} finding higher order corrections to parabolicity.

To the best of our knowledge, this has never been checked in an amorphous model of anomalous quantum Hall transition. 
To this end, we perform a detailed study at the critical point, which was extrapolated from the conductance calculation in the previous section, at $(\rho_\mathrm{c},M_\mathrm{c})=(0.7,-2.144)$. 
The distribution functions for a few moments $q$ are shown in Fig.~\ref{fig:mfa_distros} (first column). The shapes of the distribution functions become almost invariant for the larger system sizes but scale with a power law of the system size. In the second column, the horizontal axis has been rescaled, with the exponent expected from the exact parabolic prediction $\tau^{(\mathrm{p})}_q$. If the prediction is correct, we expect a collapse for large system sizes, at which irrelevant finite-size corrections no longer play a role. Indeed, for $q=-0.2,0.5$ the collapse seems to hold approximately. However, on closer quantitative inspection, we observe a residual drift of the distribution, particularly visible for $q=1.5$. The third panel shows this residual drift of the mean. Assuming perfect parabolicity with $\gamma=1/4$ and no irrelevant corrections, it should be a constant curve for all $q$. Instead, we see significant corrections, larger than the statistical error bars.

However, even though the deviations are significant, it is difficult to determine their origin conclusively:
%
%\begin{itemize}[leftmargin=*]
In particular, $q=1.5$ and $q=0.5$ show a significant curvature. This is a sign that even though the reciprocity is already converged, residual irrelevant finite-size corrections remain. To determine their exponent, larger system sizes would be necessary. Potentially, several irrelevant exponents could play a role~\cite{ObuseEPL2013}~\footnote{{
Notably, the convergence of the residual IPR has been observed in the spin quantum Hall critical point, class C for selective $q=2, 3$ values~\cite{PusHLB21}. Regardless, it has been firmly established that class C has quartic corrections in the MF spectrum~\cite{PusHLB21, KarcherAOP21}}}.

The curvature seems small {(of the order $10^{-3}$)} in particular for lower $q$'s. It might be too small to explain deviations from the collapse in the center panels. In this case, the remaining slope of the curves in the right panels is a correction to the parabolic prediction of the multifractal dimension in Ref.~\cite{Zirnbauer1999,Zir19}. This scenario would be a strong hint in opposition to the marginal scaling hypothesis, similar to the localization length exponent found in the previous sections, which is consistent with the universality of the IQHE. Deviations from parabolicity and the predicted parabolic prefactor $\gamma=1/4$ would also fully be consistent with the literature on the well-studied Chalker-Coddington networks~\cite{Evers2008parabolic,Obuse2008}.
%\end{itemize}
%%%
Assuming the latter scenario, i.e., the imperfect collapse being mainly attributed to deviations from parabolicity, it would be highly interesting to compare the resulting corrections to the multifractal dimension with the CC network results. 

%%%%
\subsubsection{Probing parabolicity: Anomalous dimension}
%%%
In Fig.~\ref{fig:parabolicity}, we show the anomalous part of the multifractal dimension as a function of the moment $q$, calculated by the numerical derivative \eqref{eq:mfa_tau} for increasing system sizes. The vertical axis has been divided by the parabolic part of the MF spectrum to highlight possible deviations. 
The data is shown without assuming a fitting function, as this is a considerable source of errors and might lead to misinterpretations regarding the presence of higher polynomial terms in the spectrum. 

There are several things we can comment on based on Fig.~\ref{fig:parabolicity}:
\begin{enumerate}[label*=(\roman*), leftmargin=*]
    \item  The effective anomalous dimension has not converged to the parabolic prediction in Eq.~\eqref{eq:deltaq} for the available system sizes. 
    \item The shape of the anomalous dimension as a function of $q$ seems stable across the shown system sizes.  
    \item The offset on the vertical axis is still shifting with system size to larger values. 
\end{enumerate}
All three of these observations are consistent with what previous studies found when studying the MF spectra of CC networks (at much higher precision)~\cite{Evers2008parabolic,Obuse2008}. 
Therefore, it may be possible that the deviations we observed in the previous section are actually due to non-parabolicity, even in the asymptotic limit. This interpretation however needs to be taken with great care since the statistical quality of data in the present model and, more importantly, the available system sizes are significantly smaller than necessary to make a definite statement. 

\paragraph*{Offset convergence:}: Another aspect of the above data is worth noting: Similarly, as in the previous sections, we observe a shift of the data to \textit{larger} values of the anomalous dimension as we increase the system size. This is in contrast to all previously studied realizations of quantum Hall criticality: For instance, in \cite{Evers2008parabolic}, the shift was in the other direction, coming from larger values of the anomalous dimension. This means that even though all previous studies, including high-precision CC data, have not been able to converge the offset of the data (in contrast to the reciprocity relation and the quartic curvature), this may enable us to give strict lower and upper bounds for the offset as well, assuming its universality across different models in the asymptotic limit. 

\begin{figure}[t]
    \centering
   \includegraphics[width=1\columnwidth]{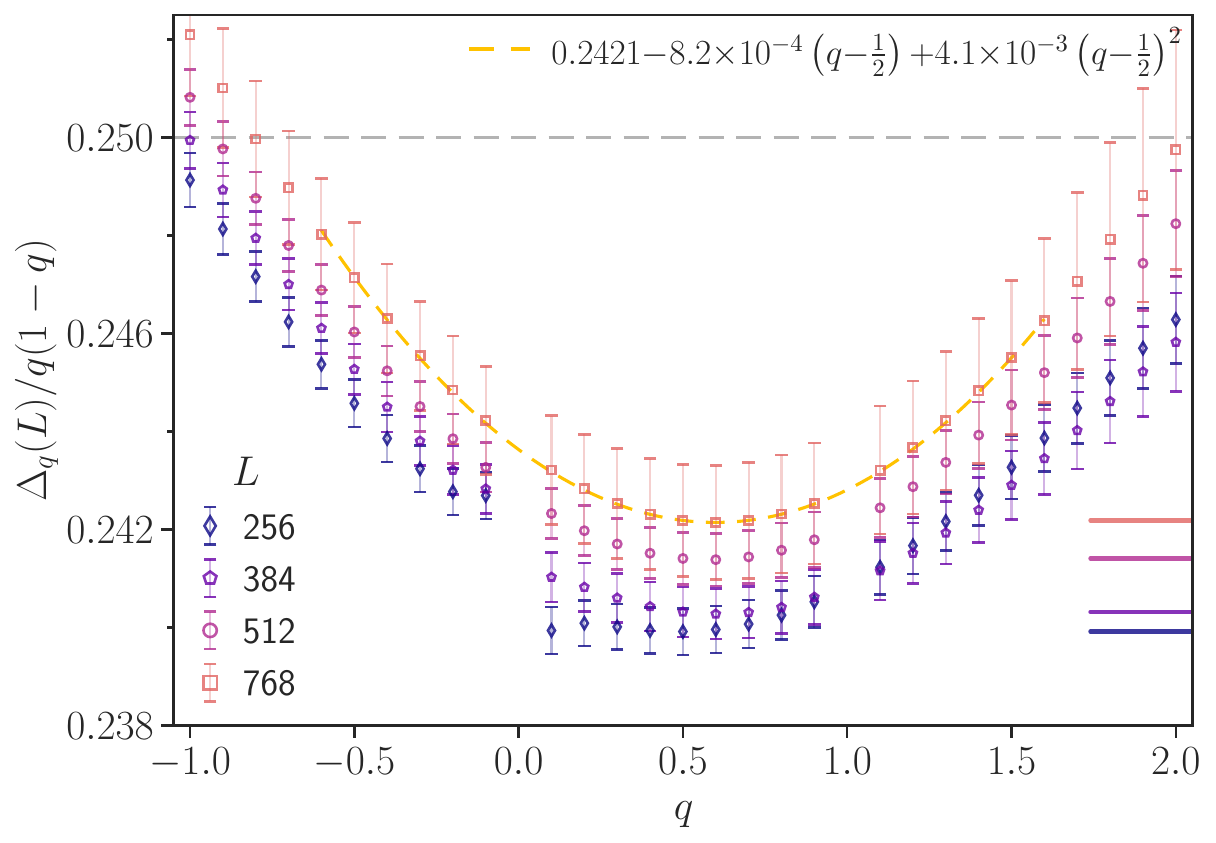}
    \caption{Highlights the deviation $\Delta_q(L)/q (1-q)$ from the parabolic multifractal spectrum~\eqref{eq:deltaq}. The data is shown without assuming any fitting function and using the pseudo numerical derivative as defined in Eq.~\eqref{eq:mfa_tau} to access the approximate $\tau_q$ directly at the critical point $\Mc\simeq -2.144$ for different combination of system sizes. 
    The dashed line is a guide to the eye with a curvature $\sim 4.1 \times 10^{-3}$. The right-hand horizontal lines indicate the flow of the minimum of these curves with system sizes.
    }
    \label{fig:parabolicity}
\end{figure}
%%%%%%%%%%%%%%%
\begin{figure}[!b]
    \centering
   \includegraphics[width=1\columnwidth]{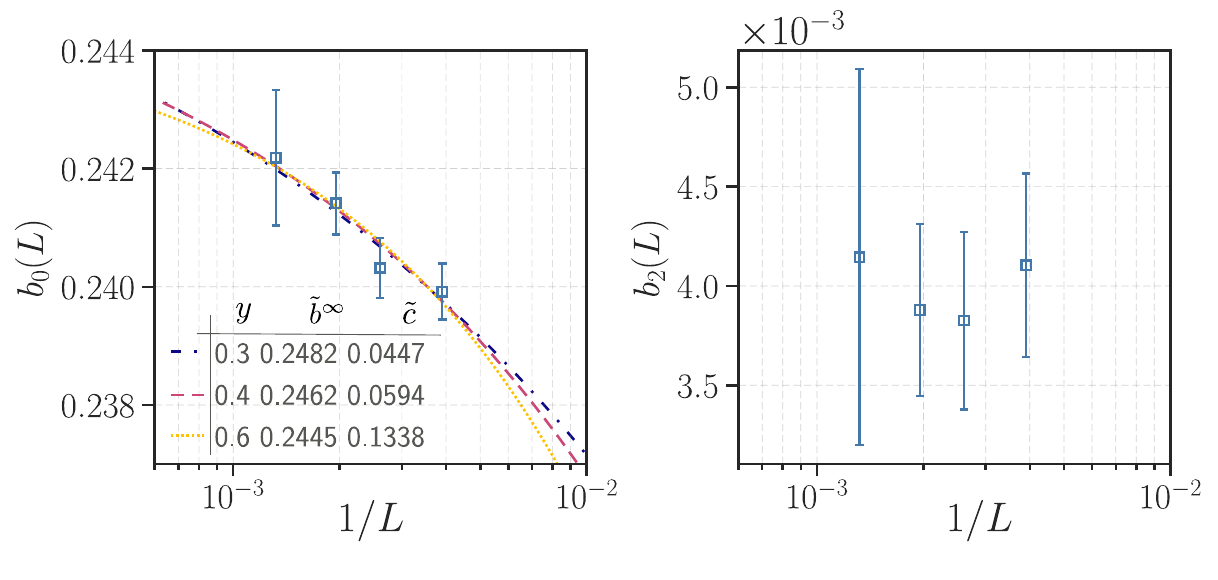}
    \caption{Show the flow of the offset $b_0(L)$, and the curvature $b_2(L)$ with system size, which is extracted from Fig.~\ref{fig:parabolicity} using Eq.~\ref{eq:residual_quartic}. Inset reports the value of the three possible extrapolations with $y=0.3, 0.4, 0.6$. (b)~Within the errorbar the curvature $b_2(L)$  shows convergence with increasing system sizes. The error bar uses the standard bootstrap procedure with $2\sigma$ confidence interval~\cite{YoungBS12}. }
    \label{fig:offset-curvature}
\end{figure}
%%%%%%%%%%%%%%%
\subsubsection{Probing parabolicity: Quartic curvature}
%%%%%%%%%%%%%%%%
Partially motivated by Ref~\cite{Puschmann23}, we characterize the residual curvature in Fig.~\ref{fig:parabolicity} as, 
\eq{
\gamma_q = b_0(L) + b_1(L) \Le(q-\f{1}{2}\Ri) + b_2(L) \Le(q-\f{1}{2}\Ri)^2 + \cdots
\label{eq:residual_quartic}
}
At the IQH critical point in the asymptotic limit, the conjecture is $b_0(L)=\gamma=\f{1}{4}$, and $b_1(L), b_2(L) \rightarrow 0$.   
In Fig.~\ref{fig:offset-curvature}, both the parameters are shown as a function of system size. 
The finite size extrapolation of the offset $b_0(L)= \tilde{b}^\infty - \tilde{c}/L^y$ is shown in the Fig.~\ref{fig:offset-curvature}(a) assuming there are different values of the irrelevant exponent $y$. 
We note that with a smaller $y=0.3$, the offset approaches $\tilde{b}^\infty\rightarrow 1/4$ ; however, given the fit window spans less than one decade, and the considerable uncertainty in the data, a smaller $\tilde{b}_\infty$ can not be excluded. 
While the offset flows with system size, the curvature within the error bars largely remains unchanged. Nonetheless, with improved statistics and system size, the flow toward vanishing offset can not be ruled out within our study. 
Current data seems to converge to $b_2 \sim 0.004(10)$, which is in the similar window with an earlier estimate $0.0058(6)$~\cite{Evers2008parabolic}.
Finally, we also observe a slight violation of the symmetry relation as the $b_1$ is finite; whether it is due to lack of statistics or has some origin could not be determined faithfully.

\section{Discussions and Conclusions}

Our work represents an essential first step in the high-precision study of the QAH critical point in an amorphous model.
While the data suggests universality, it currently doesn't match the precision of the existing data of the CC model and may not have reached the true asymptotic limit yet.
Keeping these considerations in mind, we list the main achievements of this work: 
\begin{itemize}
%[leftmargin=*, label*=(\alph*)]
%[label*={$\circ$}, leftmargin=*]
    \item The correlation length exponent is $\nu \simeq 2.60(5)$ at $E=0$.
    \item The irrelevant exponent is compatible with $y \approx 0.3(1) $.
    \item The prefactor of the irrelevant scaling $L^{-y}$ term is  smaller~(almost by a factor of 2-3) than the higher order term $L^{-2y}$, and comes with an opposite sign. 
    \item Data supports a non-parabolic MF spectrum. 
    \item The prefactor of the quadratic term converges towards its asymptotic value $\gamma=1/4$ from \emph{below} with increasing system sizes, in contrast to the CC model, where it converges from above.  
    \item The fate of the quartic curvature in the thermodynamic limit remains inconclusive. 
\end{itemize}

{\bf Exponents}: Using extensive numerical simulations, we have shown that the amorphous topological models belong to the conventional unitary ensemble and share critical properties akin to IQH criticality.
In particular, the critical exponent $\nuI$ is found to be consistent with the most recent high precision calculation on CC model $\nuI \simeq 2.609$~\cite{DresselhausPRL22}. 
However, contrary to \textcite{DresselhausPRL22}, we have used the standard finite size numerical analysis of the $\meang$ while carefully considering two subleading corrections, which turns out to be crucial. 
It is important to note that the factorization ansatz~\footnote{This ansatz of Ref.~\cite{DresselhausPRL22} proposes that the scaling function Eq.~\eqref{eq:g_expand} can be rewritten as the following:
$$
\meang(x, L) = g_1(\phi_1(x) L^{1/\nu}) g_2(\phi_2(x) L^{-y})
$$
where $x$ is the control parameter such as the $M$ or $\rho$, and $g_{1,2}$ are two independent scaling functions. Furthermore, it was argued that the $x$-dependence on the irrelevant scaling term can be ignored, i.e., $g_2(\phi_2(x) L^{-y})\equiv g_2(L^{-y})$. This would imply that the reduced conductance $\meang(x, L)/\meang(x=0, L)$ will have no irrelevant scaling corrections and is the \emph{pure} scaling variable. 
}, which has been reviewed carefully in a recent work by \textcite{KeithTomiPSS23} found to estimate consistently a smaller exponent $\nu \approx 2.55$ for the transfer matrix data~\cite{SlevinOhtsukiPRB09}. 
A similar study of the factorization ansatz in the amorphous models is performed in App.~\ref{app:alt_scale}. 
Our data confirms the $\nuI\approx 2.61$ with a similar $\chi^2$ value as observed in Fig.~\ref{fig:g_collapse}; therefore, we could not unambiguously discard the factorization ansatz against the multi-parameter fit.  
However, it should be noted that for the factorization ansatz, the critical $\Mc$ is taken from the previous analysis as shown in Fig.~\ref{fig:g_collapse}. 
We monitor a smaller $\nuI$ in the scaling of $\overline{ \log g}$ with a larger error bar.

From our current data, we can rule out an exponent smaller than approximately $\sim 2.4$. 
This finding has the following implications.
In the square lattice Chern model, recent work has demonstrated that the localization length exponent varies with energy. 
Specifically, at $E=0$ the exponent is smaller than at larger energy values~\cite{sbierskiPRL2021}. 
This result was taken as a sign of the non-universality of the anomalous quantum Hall effect, as the critical exponent depends on system parameters.
However, a thorough exploration of the irrelevant finite-size corrections in the full lattice version of the model has yet to substantiate this result. 
Instead, a recent ongoing study \textcite{Dieplinger2023} found a localization length exponent consistent with previous results $\nu\sim2.6$ at $E=0$. 
The investigations carried out in the present article extend that result to an amorphous geometry; also, here, we observe the expected IQH exponent at zero energy, thus corroborating the recent result.

Secondly, it was previously found that geometric disorder in CC model could lead to a smaller exponent~\cite{GruzbergPRB2017}. 
However, within the framework of the amorphous model, our results are inconsistent with it, though both models are significantly different. 
We consistently observe the same exponent within the error margins whether we change the density of points $\rho$ or the mass $M$, pointing towards the universality of the transition across the phase boundary.
Moreover, the observed universality contrasts with recent work on amorphous class D model~\cite{Ivaki2020PRR}, where the exponent varies across the topological transition boundary.
This variation was conjectured to result from the interplay between lattice percolation and the quantum Hall transition. 
In the model~\eqref{eq:ham} studied here, the percolation transition probably occurs at a scale significantly different than the topological transition. 
Consequently, the likelihood of observing such an interference effect is minimal; therefore, the apparent non-universality of the previous study should be taken as a model-dependent phenomenon.

In contrast to the CC model and several other studies (see, e.g., Ref.~\cite{ObuseEPL2013}), we consistently observe a slightly larger variation of the irrelevant exponent $y\approx 0.2\text{-}0.4$ for both the observables. 
The most important thing to note is that there is a conspiracy of errors. 
It implies that one would minimally require second-order irrelevant corrections, i.e., $L^{-2y}$, to see the appropriate corrections consistently, but to note that the prefactors of these corrections have opposite signs, with $c_0$ being larger than $b_0$~\eqref{eq:g_expand}. 
It further suggests that finite size corrections differ at different length scales, and one must take extreme care to distinguish that. 
A detailed analysis of this aspect would require high-precision data, which falls beyond the scope of the present work.

{\bf Conductance}: The two terminal conductance $\meang$ at the critical point is believed to be universal and characterized by a broad distribution~\cite{WangPRL96, ChoFisherPRB97, WangPRL98, WangLiSoukoulisPRB98, MarkosPRL05}. 
The experiment exhibits a near-uniform behavior, i.e., a broad distribution, within the range  $[0, e^2/h]$~\cite{CobdenKoganPRB96}. 
The amorphous model with open boundary conditions also identifies a scale-invariant broad distribution.
However, this distribution appears somewhat skewed, reminiscent of the dependence on boundary conditions on the network model~\cite{ChoFisherPRB97} or akin to observation at the 3D Anderson critical point~\cite{SlevinOhtsukiPRL00}.
In contrast, in quasi-1D geometries, the two-point conductance exhibits a log-Gaussian distribution at criticality~\cite{ObuseEPL2013}. 
Additionally, the scaling of the $\meang-\meang_\mathrm{c} \sim L^{-y}$, at the criticality agrees with an irrelevant scaling exponent $y\approx0.3(1)$, and here also, we could identify the conspiracy of errors. 
Moreover, while accessing the transition by changing the density of the points, Fig.~\ref{fig:g_collapse_with_rho}, we found a rather smaller irrelevant exponent $y\gtrsim 0.2$, which is most likely due to the lack of larger system sizes~(see Tab.~\ref{tab:cond2}), highlighting the need for larger $L$ for these analyses. 
% \SB{Atleast mention Lutken Ross, and incompatibility of it with our data.}

{\bf Multifractality}: 
The ultimate behavior of the $q$-dependence of the MF spectrum  $\Delta_q$ is challenging to ascertain, particularly in the amorphous model, due to limitations in achieving large system sizes.
Nonetheless, we can observe deviations from a truly parabolic spectrum, even concerning the QAH critical point. 
One example of this can be seen in the reduced IPR, where we notice systematic finite-size deviations from the presumed parabolic form for $\tau_q$.
Similarly, the MF spectrum obtained without any a priori assumption of the functional form exhibits deviations (though with sizable error margins) characterized by a curvature $\approx 0.0042(10)$, which is consistent with an earlier estimate $\approx 0.0058(6)$ by \textcite{Evers2008parabolic}, and by \textcite{Obuse2008}. 

\paragraph*{Flow:} 
An alternative scenario could involve the curvature $b_2(L)$ staying finite in the asymptotic limit,  while the overall deviation $b_0(L)$ converges towards $\gamma_q=1/4$.
Observably, the flow in the offset $b_0(L)$ is slow, and within our system sizes, we observe a flow exponent $y\approx 0.3(1)$, consistent with an estimate from the conductance scaling. 
However, the status of the finiteness of $b_2(L)$ in the asymptotic limit appears inconclusive in our study. 
%I

{\bf Experiments}: 
Traditionally, in experimental investigations, the critical exponents of the IQH were determined through the scaling of the transverse resistance with temperature $\Le(dR_{xy}/dB\Ri)\rvert_{B_\mathrm{c}} \propto T^{-\kappa}$, where $\kappa=p/2\nu$ and $p$ is related to the scaling of the decoherence length with temperature $L_\phi \propto T^{-p/2}$. 
Recent experimental measurements have achieved high accuracy~\cite{WeiPRL88, SaharPRL97, WanliPRL05, WanliPRL09} with $\kappa\simeq 0.42$, and even more recently at the fractional Hall plateaus~\cite{KaurKappa23}; however estimating $\nu$ requires an independent measurement of $p$.
For instance, from the coherence length measurement, \textcite{WanliPRL09} confirmed $p=2$, which would imply a localization length exponent $\nu\approx 2.38$.
This discrepancy with most numerical simulations remains unresolved.
Theoretically, under the assumption of short-range electron-electron interactions, one can estimate $p=1+2\mu_2/d$, where $\mu_2$ represents the subleading multifractal exponent~\cite{LeePRL96, WangPRB00, burmistrovAP11}.
In numerical studies in the CC model, it was determined to be $\mu_2\approx 0.62(5)$~\cite{burmistrovAP11}, implying a smaller $p\simeq 1.62$. 
In the future, conducting a similar analysis at the QAH critical point would be fruitful.

Recent experimental advancements have also focused on probing the QAH  transition within magnetic topological insulators~\cite{ChangPRL16, Liu2020, KawamuraPRB20, Wu2020, Deng2023}. 
A recent experiment by \textcite{Deng2023} measured the temperature dependence exponent in the QAH - quantum axion insulator transition and found it to be smaller $\kappa\simeq 0.34(2)$.
In a separate study, \textcite{Liu2020} reported a significantly larger value $\kappa\approx0.47$. 
Hence, it is evident that current experimental analyses exhibit significantly larger error bars when compared to both previous experiments on the IQH and numerical simulations; thus, further precise experimental studies are necessary. 

\paragraph*{Outlook:}
We conclude with a few potential directions for future research in amorphous systems. 
A recent experiment demonstrated that in 2D Dirac systems, conductance fluctuations reveal signatures of the MF spectrum at the critical point~\cite{KaziPRL22}. These measurements were close to the IQH plateau transitions in high-mobility graphene devices. Similar analyses could also be conducted in amorphous materials, and it would be exciting to probe the localization length exponent in such an experiment.

Recently, it was shown that the surface states of chiral 3D topological insulators (class AIII) at finite energies is IQH critical~\cite{SbierskiPRX20, Altland2023}. Investigating whether such states also exist in amorphous 3D models holds promise for future exploration and understanding of topological protection of critical surface states at finite energies in amorphous models. 

\section{Acknowledgements}
We thank Ferdinand Evers, Ilya Gruzberg and Martin Puschmann for several insightful discussions. We also thank Bjoern Sbierski for a critical reading of the manuscript. 
SB acknowledges support from MPG for funding through the Max Planck Partner Group at IITB. SB also thanks the MPI-PKS, Dresden computing cluster, where the conductance calculation is performed. 
JD acknowledges support from the German Research
Foundation (DFG) through the Collaborative Research
Center, Project ID 314695032 SFB 1277 (project A03)
and the German Academic Scholarship Foundation. JD also gratefully acknowledges the Regensburg computer cluster Athene, which received major instrumentation funding by the DFG, project number 464531296, and the Gauss Centre for Supercomputing e.V. and its Supercomputer SuperMUC-NG at Leibniz Supercomputing Centre through project pn36zo for computing time. 
NN would like to thank DST-INSPIRE fellowship No. IF- 190078 for funding.

\appendix

%%%%
\section{Disorder configurations}
%
%%% TABLE of the conductance
\begin{table*}[t]
\renewcommand{\arraystretch}{1.5}%
\begin{tabular}{p{20mm} | p{7mm} cccccccccc}
& $L$ & 32 & 48 & 64 & 96 & 128 & 192 & 256 & 384 & 512 & 768   \\
\midrule
 ($\meang, M$)  & $N_\mathrm{c}$ & 1.5 & 1.5 & 1.25 & 0.75 & 0.5 & 0.5 & 0.315 & 0.225 & 0.19 & 0.045 \\
($\meang, \rho$) & $N_\mathrm{c}$ &- &0.68 &0.65 &0.26 &0.20 &0.15 &0.31 &0.07 &0.06 & - \\
$(\ipr, M)\rvert_\mathrm{crit.}$ & $N_\mathrm{c}$   & 6.06 & 2.22 & 1.04 & 0.32 & 0.15 & 0.19 & 0.26 & 0.079 & 0.074  & 0.014 \\
%($\ipr, M$) & $N_\mathrm{c}$   & & & &  &  &  &  &  &   &  \\
\bottomrule               
\end{tabular}
\label{tab:dis_config}
\caption{Disorder configurations for different observables in the unit of $\times 10^6$, and for different varying parameters $M, \rho$.}
\end{table*}
The below table gives the details of the number of configurations that has been used in determining the critical properties for various observables and parameters.

%%%%%%%%%%%%%%%%
\section{Alternative Scaling}\label{app:alt_scale}
%%%%%%%%%%%%%%%%
\begin{figure}[!h]
    \centering
   \includegraphics[width=1\columnwidth]{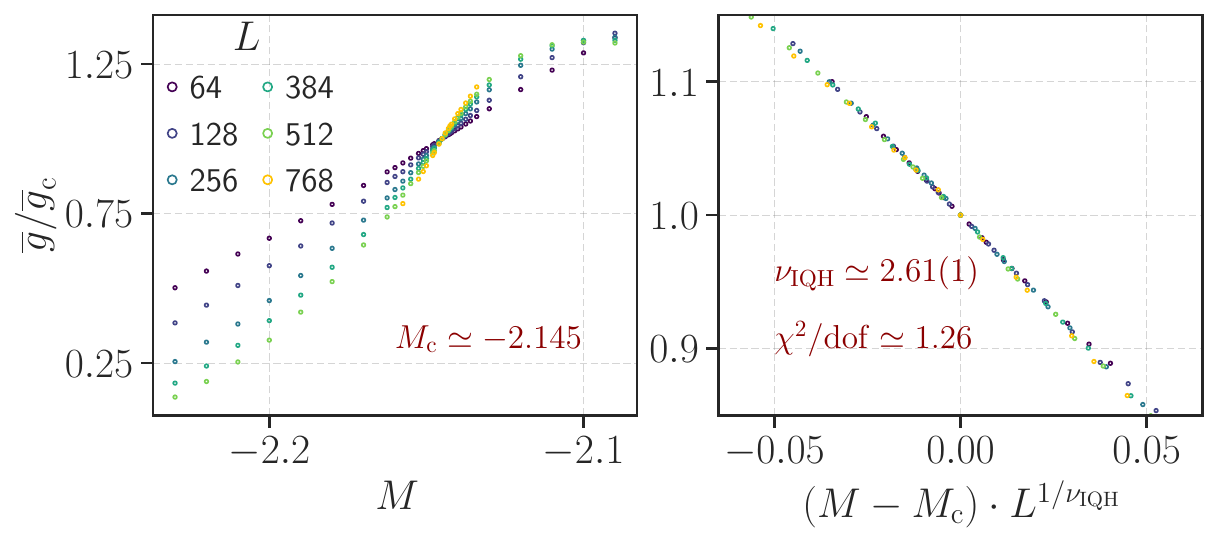}
    \caption{Showing the possible alternative scaling collapse for the same data as in the main text Fig.~\ref{fig:g_collapse}. }
    \label{fig:alt_scale}
\end{figure}
Here, we show the scaling collapse of the $\meang$ using the factorization ansatz used in Ref.~\cite{DresselhausPRL22}. 
Unlike the previous study, we required extra information of $\Mc$, which we took from the scaling ansatz fit Fig.~\ref{fig:g_collapse}, thus the analysis is not fully uncorrelated. 
Nonetheless, we observe $\nuI\simeq2.61(1)$, which is closer to the Ref.~\cite{DresselhausPRL22}. 
Importantly, the number of fitting parameters reduced drastically from 12 to 3 within the $N_\text{R}=2$ expansion order.  

\section{Fitting procedure}\label{app:fit}
%%%%%%%%%%%%%%%%
As described in the text, to achieve a stable fit, we could not keep any expansion order $N_\text{R}$. We use the Eq.~\ref{eq:g_expand}
with $N_R=2$ with two irrelevant corrections. 
This sometimes results in an unstable fit with unrealistic fitting parameters. The fitting data is provided in Tab.~\ref{tab:cond} and Tab.~\ref{tab:cond2} for different transition points. It has a lot of variability depending on the initial conditions and the number of fitting parameters. 

\begin{table*}[!htb]
\renewcommand{\arraystretch}{1.4}%
\begin{tabular}{ p{1.25cm} ccc p{1.2cm} p{1.5cm}  p{1.25cm}  p{1.25cm} p{1.25cm}  p{1.5cm} p{1.25cm}  p{1.25cm} p{1cm}}
    & & & & & \multicolumn{4}{c}{$\meang$} & \multicolumn{4}{c}{$\overline{\ln g}$} \\
    \cmidrule{6-13}
 $L$ & $N_\text{R}$ & $N_\text{I}$ & $N_\text{D}$ & $N_\text{P}$ & $\Mc$ & $\nu$ & y & $\chi^2$/dof & $\Mc$ & $\nu$ & y & $\chi^2$/dof \\ 
  \midrule
 48-768 & 2 & 2  & 170 & 12   & -2.145(1) & 2.55(2) & 0.36(3) & 1.26    & -2.144(2) & 2.69(3) & 0.61(3) & 1.14 \\
 % \rowcolor{gray!15}
 64-768 & 2 & 2  & 153 & 12   & -2.147(4) & 2.55(5) & 0.29(5) & 1.27  &-2.146(1) & 2.51(13)  &  0.37(5)  & 1.06    \\
 %96-768 & 2 & 2  & 119 & 12   & 2.64  &  0.49  & 0.97  & -2.145(4) & 2.56(5) & 0.33(20) & 1.46     \\
 \bottomrule
\label{tab:cond}
\end{tabular}

\caption{Showing the variability of the finite size scaling analysis for the transition with different initial conditions than what is shown in the main text. The $N_\text{R}$,  $N_\text{I}$,  $N_\text{D}$,  $N_\text{P}$ represent the number of relevant parameters, irrelevant expansion order, data points, and fitting parameters, respectively. }
\end{table*}
%%%

\begin{table*}[!htb]
\renewcommand{\arraystretch}{1.4}%
\begin{tabular}{ p{1.25cm} ccc p{1.5cm} p{1.25cm}  p{1.25cm}  p{1.25cm} p{1.25cm}}
    & & & &  & \multicolumn{4}{c}{$\overline{\log g}$}  \\
    \cmidrule{6-9}
 $L$ & $N_\text{R}$ & $N_\text{I}$ & $N_\text{D}$ & $N_\text{P}$ &  $\Mc$& $\nu$ & y & $\chi^2$/dof \\ 
  \midrule
 %48-512 & 2 & 2  & 170 & 11   &2.73 & 0.43&1.12    & 2.59 & 0.61 & 1.00 \\
 48-512 & 2 & 2  & 200 & 12   &0.335(7) & 2.55(9) & 0.8(4) & 4.4       \\
 64-512 & 2 & 2  & 175 & 12   &0.341(1) & 2.56(3) & 0.27(15) & 4.5      \\
 \bottomrule
\label{tab:cond2}
\end{tabular}

\caption{Variability of the finite size scaling analysis for the transition when the critical density parameter is fixed to $\rho_\mathrm{c}\simeq0.344$. The parameters are same as defined in Tab.~\ref{tab:cond}.}
\end{table*}

\section{Energy and lead dependence of the critical properties}

We examined how the critical point ($M_c=-2.144,\rho_c=0.7$) depends on the energy in Figure \ref{Critical_E_dependence}. For negative energies within a specific range, quantized plateaus emerge as the system size increases. In contrast, positive energies lead the system toward an insulating state. While we haven't conducted an exhaustive analysis of the energy dependence, it's evident that a range of energies around $E=0$ exists where the system exhibits critical behavior. Notably, in the main text, we focused on studying the critical energy at $E=0$

For the conductance simulations done in the main text, we connected a regular square lattice lead to the amorphous scattering region. For smooth connection, we embedded $L$ number of lattice sites on both sides of the scattering region~(see fig.~\ref{fig:geometry}). We analyzed the stability of the IQH critical conductance value to variations in the lead connections and lead Hamiltonian structure.
For instance, we decreased the quantity of embedded sites to $L/2$ (see fig.~\ref{fig:geometry}) and observed that the conductance value at the critical point ($g_c$) stayed constant (data not shown).
For the conductance data obtained in this work, we kept the lead hopping matrix to be $2 \times 2$ identity matrix, 
 and with zero onsite potential. The choice was made to have a finite density of lead energy states at the $E=0$. The critical conductance value remained stable to the choice of lead hopping and onsite matrix elements as long as the lead density of states remain finite at $E=0$.

\begin{figure}
    % \centering
    \includegraphics[width=1\columnwidth]{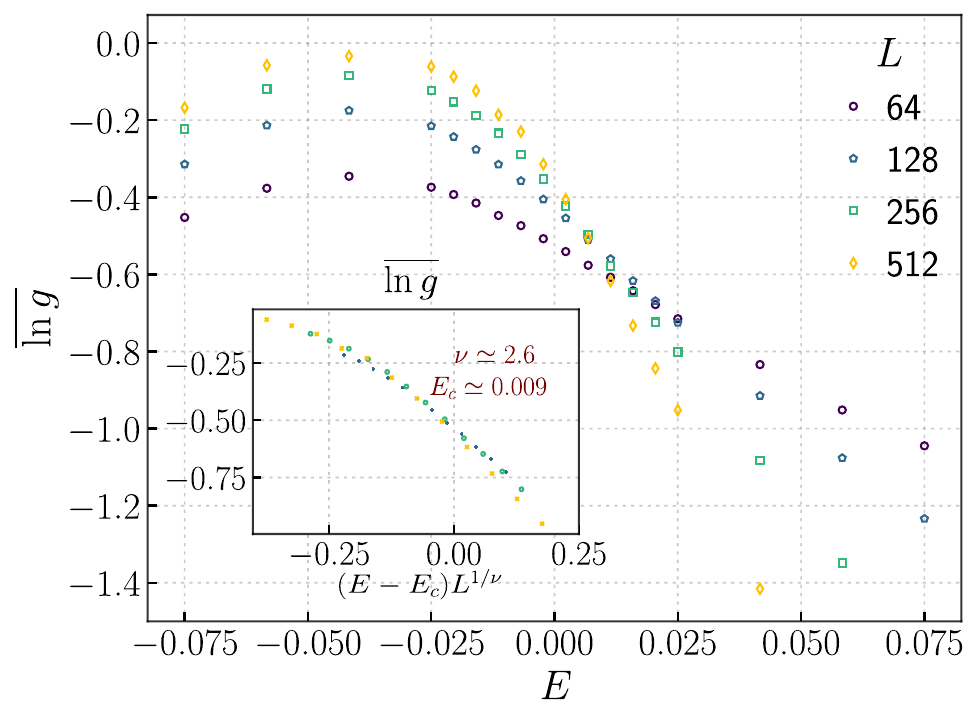}
    \caption{Energy dependence of the log-conductance at the criticality. The parameters $M=-2.144$, $\rho = 0.7$ are fixed to their critical values at $E=0$ and the energy $E$ is then varied to monitor the energy dependence. The inset figure shows the approximate scaling collapse of the various $L$ curves by rescaling the energy axis. Note here that the collapse here is just by eyeballing and no scaling function has been fitted to the data. The approximate critical parameters are shown in the figure. }
    \label{Critical_E_dependence}
\end{figure}
%%%%%%%%%%%%%%%%%%%5
\bibliography{Arxiv_version/draft_amorphous}
\end{document}